\documentclass[12pt]{article}
\usepackage{graphicx}
\begin{document}
\centerline{\bf Simulation of Rapoport's rule for latitudinal species spread}

\bigskip
Dietrich Stauffer* and Klaus Rohde**

\bigskip
\noindent
*Instituto de F\'{\i}sica, Universidade
Federal Fluminense; Av. Litor\^{a}nea s/n, Boa Viagem,
Niter\'{o}i 24210-340, RJ, Brazil; visiting from
Institute for Theoretical Physics, Cologne University, D-50923 K\"oln,
Euroland.

\bigskip
\noindent
**Zoology, University of New England, Armidale NSW 2351, Australia.

\bigskip
e-mail: krohde@une.edu.au, stauffer@thp.uni-koeln.de

\bigskip

\quad \quad Abstract: 

{\small Rapoport's rule claims that latitudinal ranges of plant and animal
species are generally smaller at low than at high latitudes. However,
doubts as to the generality of the rule have been expressed, because
studies providing evidence against the rule are more numerous than
those in support of it. In groups for which support has been
provided, the trend of increasing latitudinal ranges with latitude is
restricted to or at least most distinct at high latitudes, suggesting
that the effect may be a local phenomenon, for example the result of
glaciations. Here we test the rule using two models, a simple
one-dimensional one with a fixed number of animals expanding in a
northern or southerly direction only, and the evolutionary/ecological
Chowdhury model using birth, ageing, death, mutation, speciation,
prey-predator relations and food levels. Simulations with both models
gave results contradicting Rapoport's rule. In the first, latitudinal
ranges were roughly independent of latitude, in the second,
latitudinal ranges were greatest at low latitudes, as also shown
empirically for some well studied groups of animals.}

\bigskip
\section{Introduction}

This paper on the spread of species in North-South direction deals with
"Rapoport's rule" for land, freshwater and marine surface animals.
Stevens (1989), who named it and provided some examples, used it to
"explain" the greater species diversity in the sense that latitudinal
gradients in species diversity and Rapoport's rule have  coincidental
exceptional data and therefore have to be the "outcome of the same
process.". The rule claims that latitudinal ranges of species are
generally narrower at low than at high latitudes, which would facilitate
coexistence of more species in the tropics. It was named after Rapoport
(1982) who had earlier described this phenomenon for subspecies of
mammals. Subsequently, many papers on the rule were published, some
supporting and many providing evidence against it (there are more than
300 citations of Stevens' paper). The first papers that provided evidence
against the rule are those by Rohde (1992) and Rohde et al. (1993), using
marine teleosts (excluding deepwater and migratory species). In the same
papers, however, support for the rule was found for freshwater teleosts,
although only above a latitude of about 40 degrees North. Indeed, support
for the rule for most groups that have been examined and used as evidence
for the rule is strongest above a latitude of about 40 to 50 degrees, and
Rohde (1996) therefore concluded that the rule decribes a local
phenomenon, restricted to high latitudes. He explained this by the
extinction of species  adapted to  narrow temperature ranges during the
ice ages, an explanation also given by  Brown (1995). Another explanation
is given by the "climatic variability " or "the seasonal variability
hypothesis", first proposed by Letcher and Harvey (1994) and Stevens
(1996), according to which the greater seasonal temperature fluctuations
at high latitudes select for greater climatic tolerances and therefore
greater latitudinal ranges (for a recent discussion see Fernandez and
Vrba 2005). Stevens (1992) extended the rule to elevational gradients,
claiming that species tend to have greater altidudinal ranges toward
mountain tops, and Stevens (1996) extended the rule further to depth
gradients in the oceans. Gaston et al. (1998) concluded that support for
a general pattern described by the rule is "at the very least" equivocal.
Rohde (1999) has reviewed papers for and against the rule.

One problem with the rule is the methods used to support it. Usually,
means of the ranges in a particular latitudinal band are plotted against
latitude, although Roy et al. (1994) and Rohde and Heap (1996) used
median and modal ranges as well. Stevens (1989) counted all the species
occurring in each 5 degree latitudinal band, i.e. a species with a range
of 50 degrees appears in 10 or 11 bands. Rohde et al. (1993) have shown
that this method leads to an artificial inflation of latitudinal ranges of
high latitude species. The reason is that species diversity at low
latitudes is greater, and if only a few tropical species have large
latitudinal ranges extending into higher latitudes with their much lower
diversity, the average latitudinal range there will be greatly inflated
(the few high latitude species extending into the tropics would hardly
have any effect). They therefore proposed the midpoint method, in which
only those species are counted which have their midpoint in a particular
5 degree band. The midpoint method also has been criticised. Indeed,
there is a certain arbitrariness in using midpoints: it may not always be
the case that a species has originated in what is now its midpoint, and
so a species extending far into the tropics may in fact be an originally
high latitude species.

Also, Rapoport's rule does intuitively not make sense. The tropics have
by far the largest latitudinal range (about 23 degrees South to 23
degrees North) with more or less uniform and high temperatures. Since
temperature is one of the most important (and probably the most
important) environmental parameter affecting communities, one would
expect the largest latitudinal ranges in the tropics (as indeed shown by
Rohde et al. 1993 for marine teleosts). In the marine environment, the
trend described as "Thorson's rule" would counteract Rapoport's rule.
According to Thorson's rule, tropical marine benthic invertebrates, such
as molluscs and echinoderms, tend to produce very large numbers of small
eggs and pelagic larvae that disperse widely, whereas  high latitude
species tend to produce few and large offspring, often by viviparity or
ovoviviparity, that are often brooded, i.e. stay close to the parent
animal (Thorson 1950). That the trend is not restricted to benthic
invertebrates, was shown by Rohde (1985) who demonstrated the same
phenomenon for monogenean gill parasites of marine fishes.

Nevertheless, even application of the midpoint method sometimes shows a
Rapoport effect. The question is: why do some species have narrow and
others large latitudinal ranges in the tropics? Rohde (1998) has
suggested that newly evolved species and species with little vagility and
dispersal abilities should have narrow latitudinal ranges, and because
most species originate in the tropics, the effect would be Rapoport's
rule. In contrast, species that disperse widely would have the widest
ranges in the tropics, because uniform temperature over wide ranges
provides suitable habitats.
 
In this paper, we try to test Rapoport's rule with two different models:
A one-dimensional simple model with a fixed number of animals expanding in
northern or southern direction only, and the two-dimensional Chowdhury
model involving birth, ageing, death, mutation, speciation, prey-predator
relation and food levels (Chowdhury et al 2003, Chowdhury and Stauffer
2005, Stauffer et al 2005). The simple model is used to clarify some
definition problems for the latitudinal spread of a species. The Chowdhury
model, which earlier gave results compatible with the higher density of
animals species in the tropics (Rohde and Stauffer 2005), is then used
for a more realistic simulation. In the simple model we assume that a species
originates in a temperature region for which it is particularly fit, and
then moves to other temperatures. In contrast to simulations of Arita
(2005) we treat each animal separately.

\section{Simple model}
\subsection{Width definition}
If an animal species occurs mainly at some geographical latitude $x$  but is 
spread over an interval from $x-\Delta$ to $x+\Delta$ in North-South direction,
how do we define this width $\Delta$? 
We could determine the northernmost and the
southernmost occurrence of the species, excluding records of individuals
that have been carried by wind, currents or other external agents into
obviously abnormal habitats where the species cannot survive (e.g.,
penguins are occasionally carried to the beaches of Rio de Janeiro only
to die there). Alternatively, only nesting places or other evidence that
the species can survive and produce offspring there, could be used. The
choice will depend on the group, the question asked and the data
available. Again $2\Delta$ could be the difference between the most northern 
and the most southern latitude. However, although both definitions are
useful for the analysis of real data sets, which very rarely give
information on animal numbers at different latitudes, they are not
optimal for computer simulations and are in principle wrong (though 
unavoidable) also for reality.

Let us assume that a species centered at about the equator has a
probability of one in a million to survive at some colder northern
latitude $\Delta$. Then, if only 100,000 such animals exist, presumably none
would reach that latitude. The same effect occurs if there are a million
animals, but only ten percent are observed. If, on the other hand, the
species has 1000 million animals, and almost all those up north are
found, then the northernmost latidude observed for this species would be
larger than $\Delta$. (This criticism also applies, if all species occurring 
in some band are checked).

To avoid this dependence of the observed width on the number of animals, the
quality of observations, and the randomness of rare and extreme events, we
instead define the width $\sigma$ through the standard deviation of the
latitude of the animals:
$$ \sigma^2  = <(x - x_c)^2>$$ 
as customary in statistics. Here $x_c$ is the center latitude, often taken as 
$<x>$ and in our case defined as the latitude in the centre of distribution,
which in nature often will be the centre of origin and is treated in
the model as such.
(See below for a correction due to tunneling across the equator.) The brackets
$<\dots>$ indicate averages over all animals $i=1,2, \dots, N$, like 
$$<x> = \sum_{i=1}^N x_i \quad .$$
For a 
"normal" (Gaussian) distribution it means that about 2/3 of the animals live
in the interval between $x_c - \sigma$ and $x_c + \sigma$. If with this
definition the population changes from large to very large, then $\sigma$
is determined more accurately, without changing in a systematic way to larger 
values.  But, as pointed out above, this definition of ours is
applied much more easily to computer simulations where we can track all
animals, than to real data sets.

\subsection{Model definition}

Following Stevens, we use 35 bands of 5 degree latitude each, with the first
near the South Pole, the central band  around the equator, and the last band
near the North Pole. In each band, one typical species  of $N$ animals has 
its origin, and later these animals spread northwards and southwards to the
other bands. If an animal in the first band wants to move further south,
it is put into the second band instead. Symmetrically, an animal in the last
band 35 wanting to move north is put into band 34. The animals neither die
nor give birth and thus represent a whole sequence of generations, in the 
bad old tradition of theoretical biology of working with a constant number
of animals. The width is measured in bands, i.e. in units of 5 degrees
latitude.  

If the probability of an animal to move is constant (taken as 1/2) and  
(apart from the reflexions at the first and last band) the animals move 
North or South with equal probability, then after some time all animals 
are spread homogeneously over most of the bands, except near the extremes.
This is hardly realistic; polar bears have not expanded to the equator. 

Thus instead we use biased wandering: The probability to move in the 
direction of the band of origin is higher than for the opposite 
direction. We take this bias as $(x-x_c)/5$, where  $x = 1,2, \dots, 35$ now 
is the band number. Thus if an animal has moved five bands northwards from 
its origin, then the next move with certainty goes South.

However, this model also has to take into account the tunneling from the
northern to the southern hemisphere to the mirror latitude $-x_c$ corresponding
to the latitude of origin. With our definition of bands, the latitude is
$5(x-18)$ degrees, negative in the South and positive in the North. Then 
instead of averaging over $(x-x_c)^2$ as defined above, and of using a bias 
proportional
to $x-x_c$, we take $(x-x_c')^2$ and $x - x_c'$ whenever an animal is on the
hemisphere which is not that of its origin. Here $x'= 36-x$ is the mirror band
corresponding to the same latitude as $x$ on the other hemisphere.
In other words, the bias favours motion towards the same temperature as the 
temperature of origin, and we measure
distances only from the closer of the two temperature bands. Thus if half of
the animals live at 20 degree South, where the species has originated,
and the other half have crossed the equator and live at 20 degree North, then
the width is zero and not 20 degrees.

Finally, instead of assuming a vagility independent of latidude, we also allow
the moving probability to diminish linearly with increasing latitude, from 
unity at the equator to zero at the poles, depending on the current latitude
of the animal (justified by Thorson's rule, see above).

\subsection{Results}

\begin{figure}[hbt]
\begin{center}
\includegraphics[angle=-90,scale=0.4]{rapoport1a.eps}
\includegraphics[angle=-90,scale=0.4]{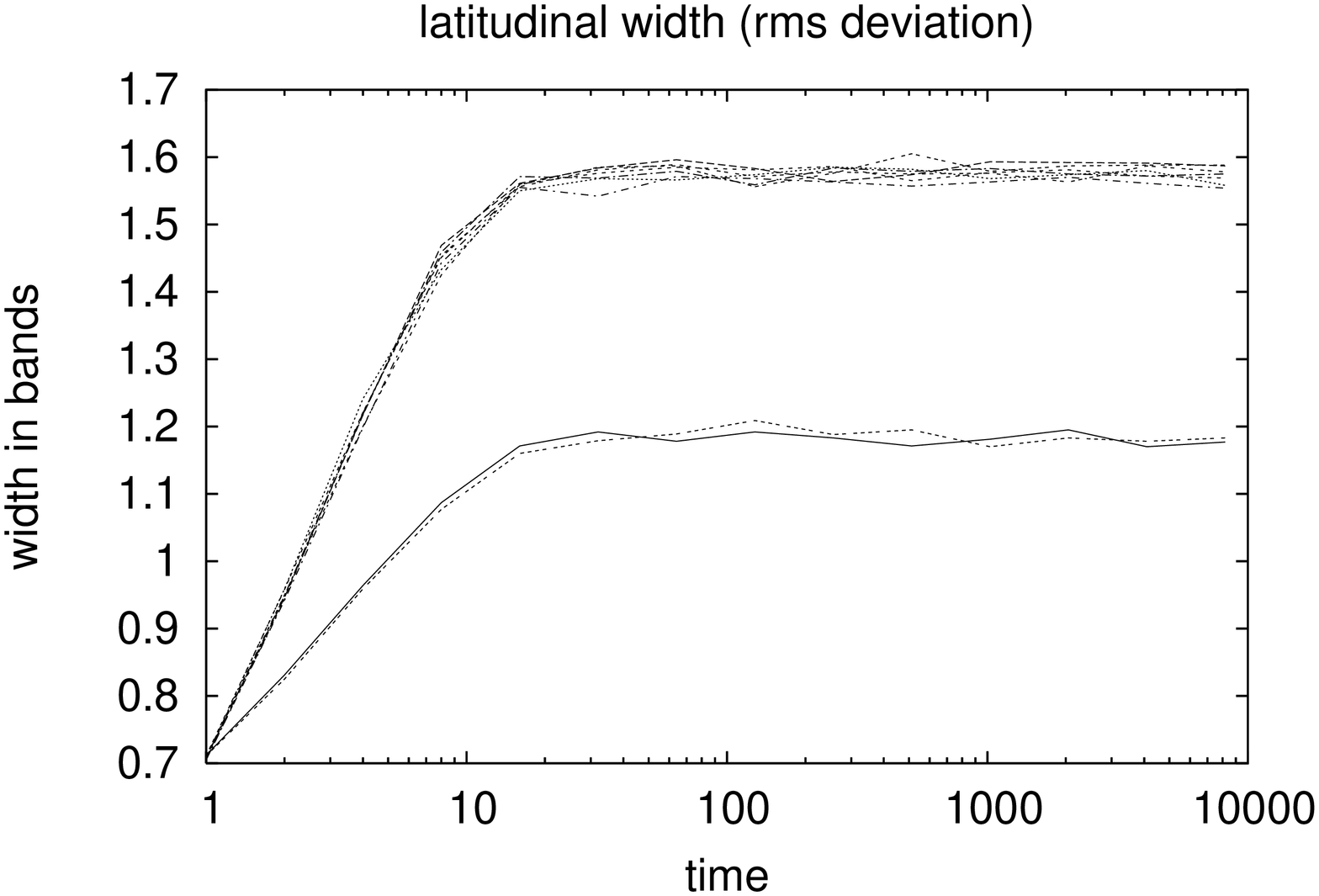}
\end{center}
\caption{Top: Distribution of animals originating at 80, 60, 40, 20, and 0 
degrees
South. We see South-North tunneling from the 20 degree band, but not from
latidude 40 and higher. The curves for animals originating in the North are
symmetric and omitted for clarity. Bottom: Time developments for the width,
including the northern species. 
}
\end{figure}

\begin{figure}[hbt]
\begin{center}
\includegraphics[angle=-90,scale=0.4]{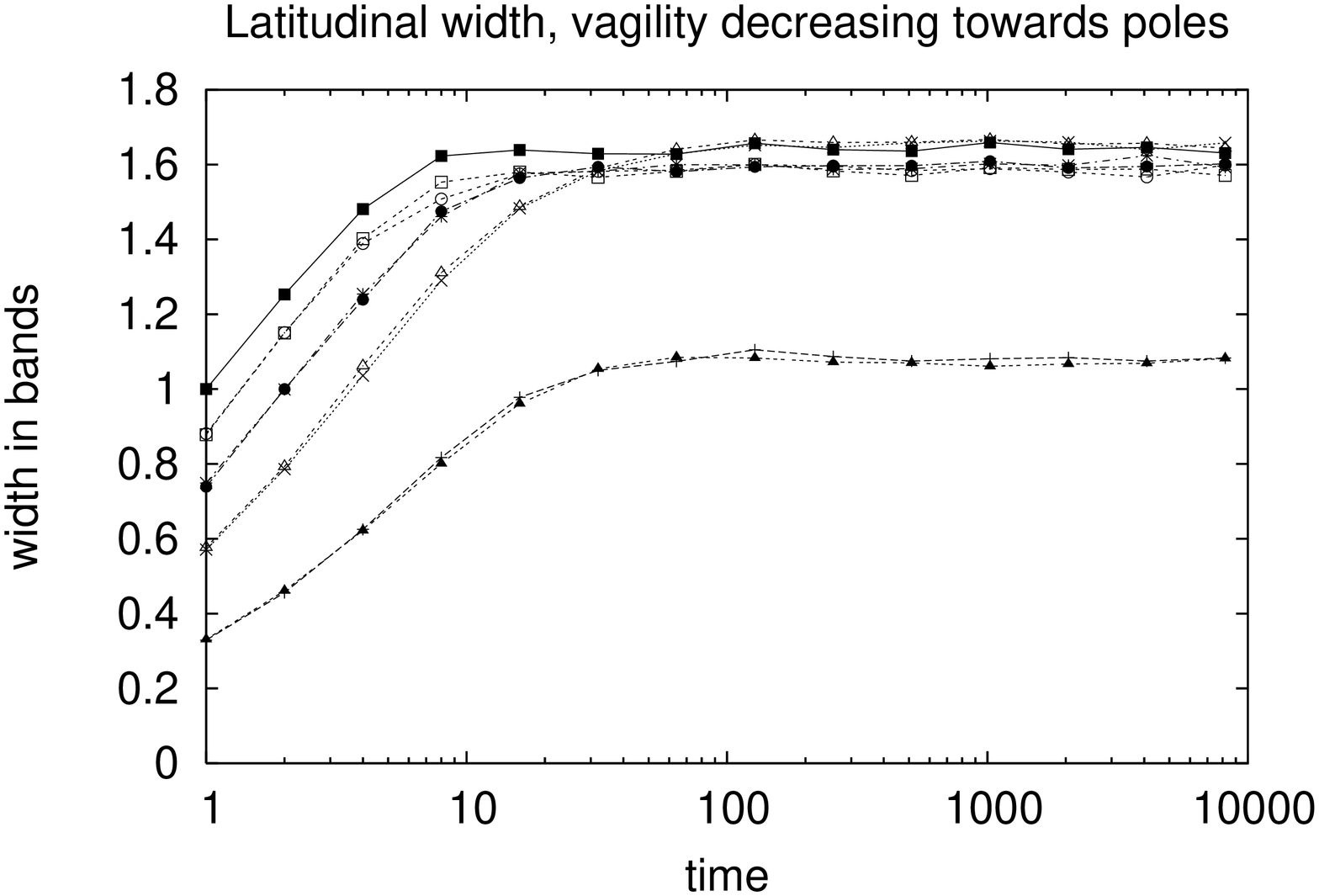}
\end{center}
\caption{As Fig.1 bottom, but with high vagility near equator (upper lines)
and low vagility near  the poles (lower lines). 
}
\end{figure}

Fig. 1 shows in its top part the animal distributions for selected bands of
origin (southern hemisphere) and in its bottom part the resulting widths
versus time. Equilibrium is seen here already after 10 to 100 time steps.
Except near the two poles, the widths are about the same, and the distributions
are Gaussian. Fig.1 uses a constant vagility, while for Fig.2 it is higher
near the equator than near the poles; in Fig.2 we now see clear differences 
in the time development of the width, but still the same equilibrium widths.
The animal distribution is about the same as in Fig.1, and is therefore not
shown in Fig.2.

\section{Chowdhury model}

\subsection{Model definition}

The simulations with the more realistic Chowdhury model, based on individual
births and deaths, is more complicated in some and simpler in other aspects. 
It represents a whole ecosystem; we took the number of food levels as six; 
tests with nine levels gave similar results. Self-organization of minimum 
reproduction age, litter size, and   
prey-predator relations is achieved through random mutations. Speciation 
happens with a probability $p$ per iteration by occupation of empty ecological 
niches through a species from an occupied niche. And each site of an $L \times
L$ square lattice carries a whole such ecosystem; a species in one ecological 
niche can invade an empty corresponding niche on a neighbouring lattice site, 
with probability $d$ (= diffusivity in physics). These 
two parameters $d$ and $p$ are varied here (mostly $d = 0.001, \; p=0.0001$); 
the other parameters are fixed
as in Rohde and Stauffer (2005). We refer to that paper or recent summaries
(Chowdhury and Stauffer 2005; Stauffer, Kunwar and Chowdhury 2005) for more 
details on the Chowdhury model. The Fortran program is available from 
stauffer@thp.uni-koeln.de (species23n.f, nearly 500 lines). It requires nearly
100 Megabytes of memory for six food levels. 

\begin{figure}[hbt]
\begin{center}
\includegraphics[angle=-90,scale=0.4]{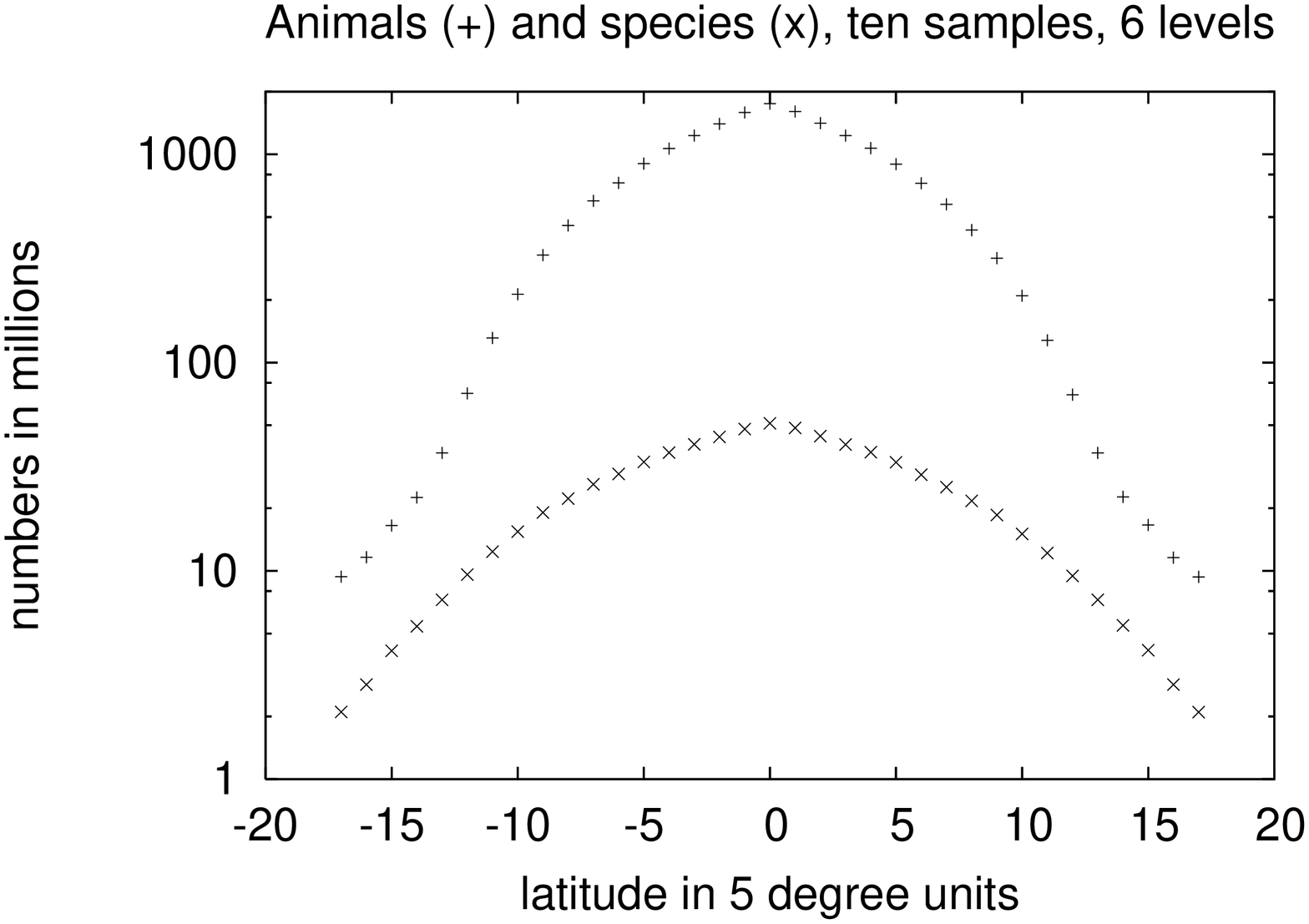}
\end{center}
\caption{Latitudinal variation of the number of animals (upper data) and 
species (lower data) in Chowdhury model, summed over ten samples run for 
240,000 time steps, of which the first 40,000 were ignored in the sums.
}
\end{figure}

\begin{figure}[hbt]
\begin{center}
\includegraphics[angle=-90,scale=0.55]{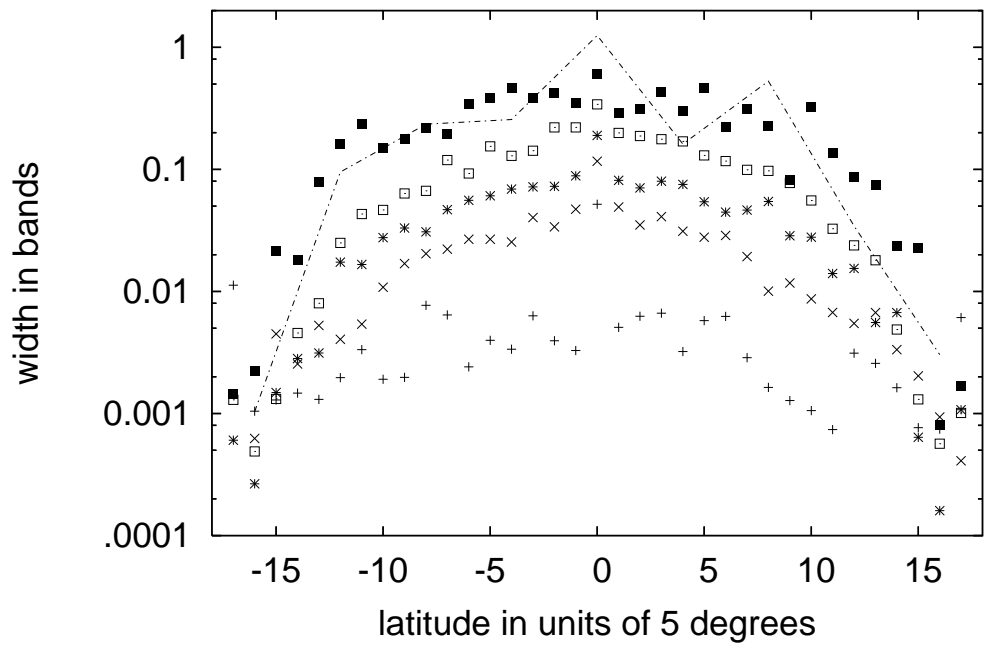}
\end{center}
\caption{Slow increase of the widths $\sigma(\ell)$, when time $t$ increases by 
factors of ten, from bottom to top. As for Fig.3
we omitted the initial sixth of the time from the averages, but increased the
time five times by a factor 10. For the longest time, shown by the line, only 
one instead of ten samples was simulated.
}
\end{figure}

\begin{figure}[hbt]
\begin{center}
\includegraphics[angle=-90,scale=0.4]{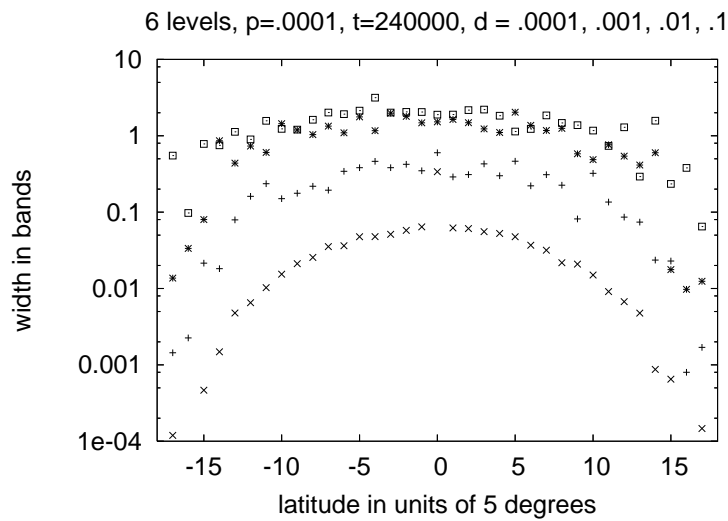}
\end{center}
\caption{Strong increase from bottom to top with increasing
diffusivity $d$ of the widths $\sigma(\ell)$; $p = 
0.0001$, statistics as for Fig.3.
}
\end{figure}

\begin{figure}[hbt]
\begin{center}
\includegraphics[angle=-90,scale=0.4]{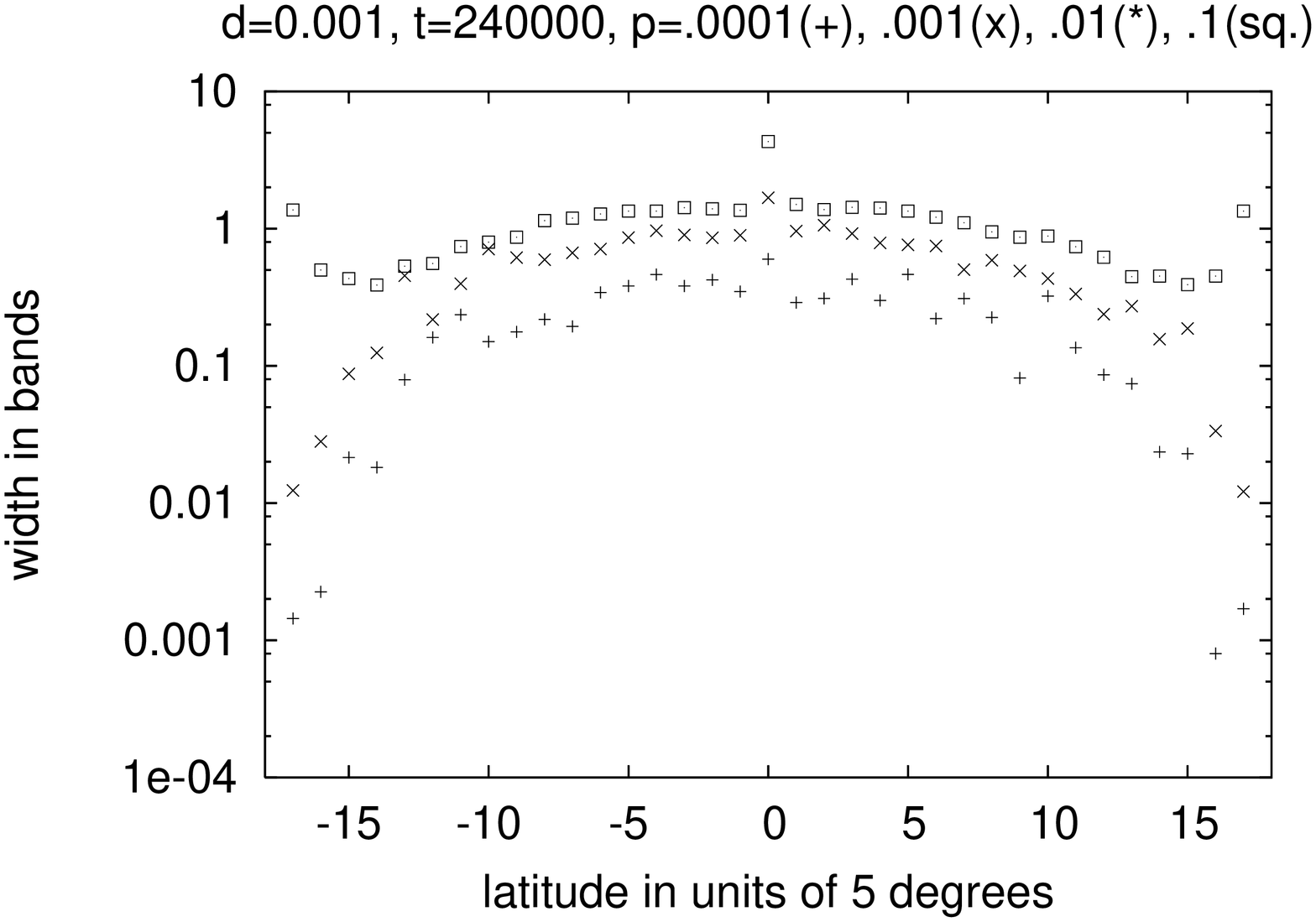}
\end{center}
\caption{Increase with speciation probability $p$ of the widths $\sigma(\ell)$;
$d = 0.001$, statistics as for Fig.3. Two higher curves still missing.
}
\end{figure}

Now we set $L = 35$ for easy comparison with the 35 latitude bands above,
and assume a birth rate proportional to $(17 - |\ell|)/17$ where $\ell = x-18$
(= latitude in 5 degree units) is the number of the band, $-17 \le\ell\le 17$.
(These birth rates count
how many animals reach maturity, and life in the cold is more difficult.) 
Again the width is measured in units of 5 degrees latitude. 

\subsection{Results}

The resulting Fig.3 shows that there are much more animals (per unit area)
near the equator than near the poles, as wanted. 
The widths $\sigma$ are again defined through $\sigma^2=\;<(\ell-\ell_c)^2> \;$
where $\ell_c$ is the band of origin of the species (not of birth of the 
individual). Figure 4 shows how these widths increase slowly (logarithmically?) 
with time $t$. Ten samples had to be averaged over to give a clear trend
larger than the statistical fluctuations. For the longest time of 2 
million (line in Fig.4) perhaps this increase stopped, but anyhow such long
times are not very realistic biologically since in a million generations 
the climate changes. The important result is that these widths are smaller
than one band, less than in the preceding section. Thus animals do not 
spread all over the world, and we did not have to introduce a bias pushing
them back to the latitude where the species originated. Thus also no 
tunneling was possible. In this sense the Chowdhury simulation was simpler than
the one of the previous section.

Fig. 5 shows that an increase of the vagility $d$ also increases the widths;
high $d$ values, however are unrealistic according to Rohde and Stauffer 
(2005). Finally, Fig.6 shows the  
variation with speciation probability $p$; again high $p$ are unrealistic.

\section{Conclusion}
ur results for both the simple model and the Chowdury model
contradict Rapoport's rule that latitudinal ranges are smallest at
low latitudes. For the simple model, ranges were more or less
independent of latitude, for the Chowdhury model, they were largest
near the equator for the parameters used. This confirms empirical
findings by various authors who did not find support for the rule in
most taxa (see above, e.g. Rohde 1993, Gaston et al. 1998). The fact
that the rule was found to hold for several groups, particularly at
high latitudes, does not contradict the results of our simulations. A
general model like ours is not meant to account for all historical
and geographical contingencies, for example those  resulting from
glaciations. If a Rapoport effect is indeed a local phenomenon due to
glaciation or other local effects, as suggested by Brown (1995) and
Rohde (1996), it would not be expected to be revealed by such a
general model. The Chowdhury model has recently been shown to give
results for latitudinal gradients in species diversity which conform
to reality (Rohde and Stauffer 2005); nevertheless, our results
concerning Rapoport's rule need confirmation from simulations using
different  models.

\bigskip
\bigskip
{\bf \large References}
\bigskip
\parindent 0pt

Arita, H.T. (2005). Range size in mid-domain models of species diversity.
Journal of Theoretical Biology 232, 119-126.
\smallskip

Brown, J. H. (1995). Macroecology. University of Chicago Press, Chicago.
\smallskip

Chowdhury, D., Stauffer D. and Kunwar A. (2003). UniÞcation of Small and
Large Time Scales for Biological Evolution: Deviations from Power Law.
Physical Review Letters 90, 068101.
\smallskip

Chowdhury, D. and Stauffer, D. (2005). Evolutionary ecology in-silico:
Does mathematical modelling help in understanding the 'generic' trends?
Journal of Biosciences (India) 30, 277-287.
\smallskip

Fernandez, M.H. and Vrba, E.S. (2005). Rapoport effect and biomic
specialization in African mammals: revisiting the climatic variability
hypothesis. Journal of Biogeography 32, 903- 918.
\smallskip

Gaston, K.J., Blackburn, T.M. and Spicer, J.I. (1998). Rapoport's rule:
time for an epitaph? Trends in Ecology and Evolution 13, 70-74.
\smallskip

Letcher, A. J., and P. H. Harvey. (1994). Variation in geographical range
size among mammals of the Palearctic. American Naturalist 144, 30-42.
\smallskip

Neumann, G. and Pierson, W.J.(1966). Principles of physical oceanography.
Prentice-Hall, Englewood Cliffs, N.J.
\smallskip

Rapoport, E.H. (1982). Areography. Geographical strategies of species.
Pergamon Press, New York.
\smallskip

Rohde K. (1985). Increased viviparity of marine parasites at high
latitudes. Hydrobiologia 127, 197-201.
\smallskip

Rohde, K. (1992). Latitudinal gradients in species  diversity: the search
for the primary cause. Oikos 65, 514-527.
\smallskip

Rohde, K., Heap, M. and Heap, D. (1993). Rapoport's rule does not apply
to marine teleosts and cannot explain latitudinal gradients in species
richness.  American Naturalist, 142, 1-16.
\smallskip

Rohde, K. (1996). Rapoport's Rule is a local phenomenon and cannot
explain latitudinal gradients in species diversity. Biodiversity Letters,
 3, 10-13.
\smallskip

Rohde, K. and Heap, M. (1996). Latitudinal ranges of teleost fish  in 
the Atlantic and Indo-Pacific Oceans. American Naturalist 147, 659-665.
\smallskip

Rohde, K. (1998). Latitudinal gradients in species diversity. Area
matters, but how much? Oikos 82, 184-190.
\smallskip

Rohde, K. (1999). Latitudinal gradients in species diversity and 
Rapoport's rule revisited: a review of recent work, and what can 
parasites teach us about the causes of the gradients? Ecography, 
22, 593-613
\smallskip

Rohde, K. and Stauffer, D. (2005). Simulation of geographical trends in
Chowdhury ecosystem model, e-print q-bio.PE/0505016 at www.arXiv.org.
\smallskip

Roy, K., Jablonski, D. and Valentine, J.W. (1994). Eastern Pacific
molluscan provinces and latitudinal diversity gradients: no evidence for
Rapoport's rule. Proceedings of the National Academy of Sciences of the
USA 91, 8871-8874
\smallskip

Stauffer, D., Kunwar A. and Chowdhury D. (2005). Evolutionary ecology in
-silico: evolving foodwebs, migrating population and speciation. 
Physica A 352, 202-215.
\smallskip

Stevens, G.C. (1989). The latitudinal gradients in geographical range:
how so
many species co-exist in the tropics. American Naturalist 133, 240-256.
\smallskip

Stevens, G.S. (1992). The elevational gradient in altitudinal range: an
extension of Rapoport's latitudinal rule to altitude. American Naturalist
140, 893-911.
\smallskip

Stevens, G.C. (1996). 1996. Extending Rapoport's rule to Pacific marine
fishes. Journal of Biogeography 23, 149-154.
\smallskip

Thorson, G. (1950). Reproductive and larval ecology of marine bottom
invertebrates. Biological Reviews 25, 1-45.
\smallskip
\end{document}